\documentclass[aps,pra,reprint,groupedaddress,nofootinbib]{revtex4-1}

\usepackage{amsmath,amssymb,graphicx}
\usepackage{array}
\usepackage[caption=false]{subfig}
\usepackage{bm}
\usepackage{amsfonts}
\usepackage{epsfig,epstopdf}
\usepackage{subfig}
\usepackage{url}
\usepackage[utf8x]{inputenc}
\usepackage[T1]{fontenc}

\def\Tr{{\rm Tr}}

\def\vR{\vec R}

\def\la{\left\langle}
\def\ra{\right\rangle}

\def\br{{\bf r}}

\def\bu{{\bf u}}

\def\hrho{\hat \rho}

\def\bH{{\bf H}}

\def\be{\begin{equation}}
\def\ee{\end{equation}}
\def\ba{\begin{align}}
\def\nn{\nonumber\\}

\def\defeq{\buildrel \rm def \over =}
\def\bR{{\bf R}}

\def\b0{{\bf 0}}

\def\la{\left\langle}
\def\ra{\right\rangle}

\def\br{{\bf r}}

\def\bu{{\bf u}}

\def\bH{{\bf H}}

\def\bl{\bm{l}}
\def\blperp{{\bm{l}}_\perp}
\def\brho{{\boldsymbol\rho}\,}

\def\Tr{{\rm Tr\ }}

\def\be{\begin{equation}}
\def\ee{\end{equation}}
\def\ba{\begin{align}}
\def\nn{\nonumber\\}
\def\la{\langle}
\def\ra{\rangle}

\def\defeq{\buildrel \rm def \over =}
\def\bR{{\bf R}}
\def\Re{{\rm Re}\,}
\def\Im{{\rm Im}\,}
\def\rH{{\rm H}}
\def\pmu{\partial_\mu}
\def\pnu{\partial_\nu}
\def\px{\partial_x}
\def\py{\partial_y}
\def\pz{\partial_z}
\def\th{\tilde{h}}
\def\tg{\tilde{g}}

\begin{document}

\title{Quantum limited super-resolution of an unequal-brightness source pair in three dimensions}

\author{Sudhakar Prasad}
\affiliation{School of Physics and Astronomy, University of Minnesota, Minneapolis, MN 55455}
\email{prasa132@umn.edu}
\altaffiliation{also at Hennepin Healthcare Research Institute, Minneapolis, MN 55404}

\date{\today}


\begin{abstract}

This paper extends our previous quantum Fisher information (QFI) based analysis of the problem of separating a pair of equal-brightness incoherent point sources in three dimensions to the case of a pair of sources that are unequally bright. When the pair's geometric center is perfectly known in advance, QFI with respect to the estimation of the three separation coordinates remains independent of the degree of brightness asymmetry. For the experimentally more relevant case of perfect prior knowledge of the pair's brightness centroid, however, such QFI becomes dependent on the pair separation vector in a way that is controlled by the degree of its brightness asymmetry. This study yields potentially useful insights into the analysis of a more general superresolution imaging problem involving extended incoherent sources with nontrivial brightness distributions.
\end{abstract}

\vspace{-1cm}


\maketitle
\section{Introduction}\label{sec:intro}
In the photon counting limit, wavefront projections via spatial demultiplexing (SPADE), as proposed recently by Tsang {\it et al.} \cite{Tsang16,Ang17}, can overcome the celebrated Rayleigh limit \cite{Rayleigh1879} for resolving a symmetrical pair of incoherent point sources  at photon numbers that scale inverse-quadratically with separation even at the smallest separations.  This scaling law, which was first derived in Refs.~\cite{Tsang16,Ang17} by applying quantum estimation theory to the problem of estimating the pair separation in one and two dimensions, seems to be entirely independent of the spatial dimensionality of the problem, as we showed in two papers \cite{YuPrasad18,PrasadYu19} via an analysis considerably more general than that of Ref.~\cite{Napoli19} for full three-dimensional (3D) source pair separation and localization using apertures of arbitrary shape and size. It amounts to a qualitatively more modest requirement than that entailed by intensity based detections of the source pair on an image sensor for which this scaling law, for small separations, is no better than inverse quartic in separation.  Extensions to thermal source states \cite{NairTsang16} and general quantum source states \cite{LupoPirandola16} have also been considered. A number of recent experiments have confirmed the achievability of the quantum estimation bound both for lateral \cite{Paur16,Yang16,Steinberg17,Parniak18} and axial \cite{Boyd19} resolutions. 

The dramatic difference in the scaling behaviors of the sensitivity of wavefront and intensity based methods opens a novel, highly sensitive approach of superresolution imaging of relatively faint extended sources for which image intensity based methods are likely to fail. More recently, SPADE has been analyzed as a potentially sensitive approach for optically acquiring an extended luminous object that is too faint and small to be resolved and imaged by ordinary intensity based methods. The method is essentially based on acquiring first, second, and higher-order intensity moments of an extended incoherent object \cite{Tsang17, Chrostowski17, Tsang19, Zhou19} using a suitable projection basis. Successful acquisition of such moments can enable one to characterize the object in terms of its brightness attributes such as centroid, size, ellipticity, skewness, etc. These methods can be potentially generalized further to extended 3D sources.  

An alternative approach for imaging an extended incoherent source might regard it as a collection of point sources \cite{Dutton19} with spatially varying intensity across it. The simplest example of such an extended object with a nonuniform brightness distribution is a pair of incoherent point sources of unequal intensity. We consider in this paper the problem of quantum limited superresolution of such a source pair.  We will use the concepts of quantum Fisher information (QFI) and quantum Cram\'er Rao bound (QCRB) from quantum estimation theory put forth by Helstrom \cite{Helstrom76} to evaluate the fidelity of estimation of the separation vector for an arbitrary ratio of the intensities of the two point sources. The problem of calculating such a quantum limit on simultaneously estimating the geometric center, separation, and relative intensity of an unequally bright source pair using these concepts was first tackled by $\check{\rm R}$eh\'a$\check{\rm c}$ek, {\it et al.} \cite{Rehacek17, Rehacek18}, but their work was limited to one spatial dimension. 

In the present paper, we will generalize the work by $\check{\rm R}$eh\'a$\check{\rm c}$ek's group, as well as our previous work \cite{YuPrasad18} to an asymmetrically bright source pair in all three dimensions. Rather astoundingly, as we will show, the QFI whose independence from pair separation is the basis of such scaling remains independent of this separation even for an unequally bright source pair when its geometric center is well determined and fixed {\it a priori}. But for the experimentally more relevant scenario where only the brightness centroid, rather than the geometric center, can be reliably determined in advance, such independence obtains only for the equal-brightness case for which the two centroids coincide. We derive here the detailed dependence of QFI on the pair brightness ratio and separation for estimating the latter under these conditions.

Let us consider a pair of closely spaced incoherent point sources that have unequal average intensities $I_\pm$, so the quantum state of a photon emitted by such a source pair is described by the density operator (DO),
\be
\label{rho}
\hrho =p_+| K_+\ra\la K_+|+p_-|K_-\ra\la K_-|,\ \ p_\pm\defeq {I_\pm\over I_++I_-},
\ee
in which $p_\pm$ may be regarded as the probability of the photon being emitted by the source in the state $|K_\pm\ra$, respectively. We consider only those photons that are captured and detected by the observing instrument, as only they can encode information about the source pair that is recoverable by means of measurements made on them. We further restrict attention to the classically ideal regime of photon counting without any detection noise under which the photons detected during a fixed observation time have a Poisson number distribution. In this regime, which applies to sources of small intensity fluctuations per mode such as ordinary thermal sources possessing a small degeneracy parameter, we may regard the photons as arriving in a statistically independent fashion at the instrument sensor. Consequently the Fisher information (FI) for the problem of $N$ photocounts, whether classical or quantum, becomes simply $N$ times the FI per photon. It is the quantum version of this FI per photon, namely QFI, that we set out to calculate. The diagonal elements of the inverse of the QFI matrix with respect to (w.r.t.) a set of $n$ system parameters, $\{\theta_1,\ldots,\theta_n\}$, on which $\hrho$ depends determine the lowest possible variances for an unbiased estimation of these parameters, regardless of any specific quantum measurement, and depend only on the quantum state of the source photons.

\section{Quantum Limited Estimation of Source-Pair Separation for Known Geometric Center and Relative Brightness}\label{sec:problem}
In the first half of the paper, we take the two point sources to be located at $\pm \bl$, relative to their geometric center chosen to be the origin. Subsequently we shall refer their locations relative to the intensity centroid, a point that is likely to be easier to locate based on image intensity measurements on a pixelated sensor array when the sources are close together. 
\subsection{Photon Wavefunction}  

As we have noted previously \cite{YuPrasad18}, the corresponding normalized wavefunctions take the following form over the exit pupil:
\be
\label{wavefunction}
\la \bu| K_\pm\ra = \exp(\pm i\phi)\, P(\bu)\, \exp[\mp i\Psi(\bu;\bl)],
\ee
in which $P(\bu)$ is a generally-complex pupil function obeying the normalization condition over the pupil plane,
\be
\label{norm}
\int d^2u\, |P(\bu)|^2 = 1,
\ee 
$\bu$ is the normalized pupil-plane position vector, which for a circular pupil is obtained by dividing the physical pupil-plane position vector by the radius of the pupil, and the phase function, $\Psi(\bu;\bl)$, for a low numerical-aperture system \cite{Goodman96} has the form,
\be
\label{pairphase}
\Psi(\bu;\bl)=2\pi \bu\cdot\blperp +\pi u^2l_z,   
\ee
which is linear in the vector of 3D spatial-separation parameters, $\bl\defeq (\blperp,l_z)$. We adopt a convenient convention of writing 3D spatial vectors in terms of their 2D transverse projections, such as $\blperp$, and axial coordinates, such as $l_z$. The actual physical 3D separation coordinates are proportional to these parameters via two different, transverse and axial, diffraction scales, as \cite{PrasadYu19}
\be
\label{ScaledSeparation}
\bR_\perp={\lambda z_I\over R} \blperp;\ \ Z={\lambda z_O^2\over R^2} l_z,
\ee
in which $(\bR_\perp,Z)$ is half the physical 3D vector separating the two sources, $z_O,z_I$ are the average object-plane and image plane distances from the exit pupil of the imager, $\lambda$ is the optical wavelength, and $R$, for a circular pupil, is the radius of the exit pupil. For a  more general pupil shape, $R$ could be regarded as a characteristic size of the pupil. For a thin-lens imager \cite{Goodman96}, the extrance and exit pupils coincide with the lens aperture.
 
The phase constant, $\phi$, is conveniently chosen to make the inner product, $\Delta\defeq \la K_-|K_+\ra$, real and positive, which implies two equivalent relations for $\Delta$,
\be
\label{Delta0}
\Delta =\la K_-|K_+\ra=\la K_+|K_-\ra.
\ee
 In view of relations (\ref{wavefunction}) and (\ref{pairphase}) for the wavefunction and $\Psi$, this inner product may be expressed as
\ba
\label{Delta}
\Delta \defeq \la K_+|K_-\ra=&\exp(-2i\phi)\int d^2u\, |P({\bf u})|^2\nn
&\times \exp(i4\pi{\bf l}_\perp\cdot{\bf u}+i2\pi l_z u^2).
\end{align}
Note that the phase constant, $\phi$, being half of the complex phase of the integral on the right-hand side (RHS) of Eq.~(\ref{Delta}), depends, in general, on the pair separation vector, $\bl$. 
For the clear, unit-radius circular aperture, the pupil function $P(\bu)$ is simply $1/\sqrt{\pi}$ times the indicator function for the aperture, but our calculations make no reference to the actual form of the pupil function, which is subject only to the normalization condition (\ref{norm}).

\subsection{Pair-Separation QFI}

The QFI matrix, $\bH$, is defined \cite{Helstrom68,Helstrom70,Helstrom76,Braunstein94} to have elements ${\rm H}_{\mu\nu}\defeq \Re \Tr(\hrho \hat L_\mu \hat L_\nu)$, where $\Re$ denotes the real part and $\hat L_\mu$ is the SLD of the DO w.r.t. the $\mu$th parameter. The QFI matrix elements may be readily shown \cite{YuPrasad18} to have the form,
\be
\label{QFI}
\rH_{\mu\nu}=\Re H_{\mu\nu},
\ee
in which $H_{\mu\nu}$ may be expressed as
\begin{align}
\label{cQFI}
H_{\mu\nu}&=\sum_{i=\pm}{4\over e_i}\langle e_i|\partial_\mu \hat\rho\partial_\nu \hat\rho|e_i\rangle\nn
&+\sum_{i=\pm}\sum_{j=\pm}\left[{4e_i\over {(e_i+e_j)}^2}-{4\over e_i}\right]\langle e_i|\partial_\mu \hat\rho|e_j\rangle\langle e_j|\partial_\nu \hat\rho|e_i\rangle.
\end{align}
Here $e_\pm$ and $|e_\pm\ra$ are the two non-vanishing eigenvalues and the corresponding eigenstates of DO (\ref{rho}), respectively, and $\partial_\mu\defeq \partial/\partial \theta_\mu$ denotes the first order partial derivative w.r.t. the parameter $\theta_\mu$, of the quantity that immediately  follows it.\footnote{We avoid the use of parentheses to ease notation except when the derivative is to be taken of multiple quantities which we enclose in parentheses.} By decomposing the double sum in Eq.~(\ref{cQFI}) into its diagonal $(i=j)$ and off-diagonal ($i\neq j)$ terms and noting that (see \cite{YuPrasad18}, Supplement)
\be
\label{diagME}
\la e_i|\partial_\mu \hrho|e_i\ra =\partial_\mu e_i
\ee
and the DO trace norm relation,
\be
\label{eigensum}
e_++e_-=1, 
\ee
we may express Eq.~(\ref{cQFI}) as
\begin{align}
\label{ccQFI}
H_{\mu\nu}&=\sum_{i=\pm}{4\over e_i}\langle e_i|\partial_\mu \hrho\partial_\nu \hrho|e_i\rangle
-{3\over e_+ e_-}\partial_\mu e_+\partial_\nu e_+\nn
&-4\sum_{i\neq j\atop i,j\in \{+,-\}}\left({1\over e_i}-e_i\right)\langle e_i|\partial_\mu \hrho|e_j\rangle\langle e_j|\partial_\nu \hrho|e_i\rangle.
\end{align}

The range space of $\hrho$ is spanned by the two states $|K_\pm\ra$ in which we may find its eigenstates as the linear combinations,
\be
\label{eigenstates}
|e_\pm\ra =\alpha_\pm |K_+\ra +\beta_\pm |K_-\ra.
\ee
The coefficients, $\alpha_\pm,\beta_\pm$, as well as the eignevalues $e_\pm$, may be determined by substituting expressions (\ref{eigenstates}) and (\ref{rho}) into the eigen-relation,
\be
\label{eigenrelation}
\hrho |e_\pm \ra=e_\pm|e_\pm\ra,
\ee
comparing coefficients of the two linearly independent states $|K_\pm\ra$ on both sides of this equation, and requiring that the determinant of the underlying homogeneous linear system of equations in these coefficients vanish, we arrive, after some algebraic simplifications, at the following expressions for $e_\pm$ and coefficients $\alpha_\pm,\beta_\pm$:
\ba
\label{eigenvalues_coeff}
e_\pm =&{1\over 2}(1\pm \delta e), \ \delta e\defeq \sqrt{\delta p^2+\Delta^2(1-\delta p^2)},\nn
\alpha_\pm=&\left[{p_+(\delta e\pm \delta p)\over \delta e(1\pm \delta e)}\right]^{1/2},\ 
\beta_\pm=\pm\left[{p_-(\delta e\mp \delta p)\over \delta e(1\pm \delta e)}\right]^{1/2},
\end{align}
in which $\delta e=e_+-e_-$ and $\delta p = p_+-p_-$ are the differences between the two eigenvalues and the two state probabilities, respectively. Note that since $\Delta^2,\delta p^2 \leq 1$, it follows from Eq.~(\ref{eigenvalues_coeff}) that $|\delta p|\leq \delta e \leq 1$. Equivalently, $e_+\geq \max(p_+,p_-)\geq  e_-\geq 0$. The unit normalization of the eigenstates, namely
\be
\label{norm2}
\la e_\pm |e_\pm\ra=1,
\ee
which in view of eigenstate expansion (\ref{eigenstates}) and definition (\ref{Delta}) of $\Delta$ is equivalent to the requirement,
\be
\label{norm3}
\alpha_\pm^2+\beta_\pm^2+2\alpha_\pm\beta_\pm \Delta =1,
\ee
was used to fix the overall normalization of expressions (\ref{eigenvalues_coeff}) for the two pairs of coefficients.

One may easily evaluate the partial derivative of expression (\ref{eigenvalues_coeff}) for the eigenvalues w.r.t. a parameter $\theta_\mu$ as
\be
\label{d_eigenvalues}
\partial_\mu e_\pm =\pm {\Delta (1-\delta p^2)\over 2\delta e}\partial_\mu \Delta,
\ee
which allows the second term on the RHS of Eq.~(\ref{ccQFI}) to be evaluated in terms of a product of derivatives of the state-overlap function, $\Delta$. We may thus simplify Eq.~(\ref{ccQFI}) as 
\ba
\label{cccQFI}
H_{\mu\nu}&=\sum_{i=\pm}{4\over e_i}\langle e_i|\partial_\mu \hrho\partial_\nu \hrho|e_i\rangle
-{3\Delta^2(1-\delta p^2)^2\over \delta e^2(1-\delta e^2)}\partial_\mu \Delta\partial_\nu \Delta\nn
&-4\sum_{i\neq j\atop {i,j\in\{+,-\}}}\left({1\over e_i}-e_i\right)\langle e_i|\partial_\mu \hrho|e_j\rangle\langle e_j|\partial_\nu \hrho|e_i\rangle.
\end{align}

By using expression (\ref{rho}) for DO in terms of the corresponding derivatives of the single-source emission states, $|K_\pm\ra$, we may evaluate its derivatives and their bilinear products as
\ba
\label{prho}
\partial_\mu\hrho =&p_+\left(\pmu|K_+\ra\la K_+|+|K_+\ra\pmu\la K_+|\right)\nn
&+p_-\left(\pnu|K_-\ra\la K_-|+|K_-\ra\pnu\la K_-|\right);\nn
\partial_\mu\hrho\partial_\nu\hrho = &\Big[p_+^2\Big(\pmu|K_+\ra\la K_+|\pnu| K_+\ra\la K_+|+\pmu|K_+\ra \pnu\la K_+|\nn
+|K_+\ra\pmu&\la K_+| \pnu| K_+\ra\la K_+|
+|K_+\ra\pmu\la K_+| K_+\ra\pnu\la K_+|\Big)\nn
+p_+p_-\Big(&\pmu|K_+\ra\la K_+|\pnu|K_-\ra\la K_-|
+\pmu|K_+\ra \Delta \pnu\la K_-|\nn
+|K_+\ra\pmu&\la K_+| \pnu| K_-\ra\la K_-|
+|K_+\ra\pmu\la K_+| K_-\ra\pnu\la K_-|\Big)\nn
+&\ {\rm terms\ obtained\ by}\  +\leftrightarrow - \ {\rm interchange}\big].
\end{align}
Substituting expressions (\ref{prho}) and expansions (\ref{eigenstates}) of the eigenstates $|e_\pm\ra$ into Eq.~(\ref{cccQFI}), we can now straightforwardly calculate its sums, which involve the matrix elements $\la e_\pm |\partial_\mu \hrho\partial_\nu \hrho|e_\pm\ra$ and $\la e_+|\partial_\mu \hrho|e_-\ra$. Taking the real part of $H_{\mu\nu}$ thus obtained yields the $\mu\nu^{\rm th}$ matrix element of QFI. This process is greatly simplified by noting from the form of the wavefunctions (\ref{wavefunction}) that matrix elements like $\la K_\pm|\pmu|K_\pm\ra$ are purely imaginary and those like $\pmu\la K_\pm|\pnu|K_\pm\ra$ purely real, since each partial derivative of either wavefunction w.r.t. a separation coordinate $l_\mu, \mu=1,2,3$, generates a purely imaginary factor. Further, by differentiating Eqs.~(\ref{Delta0}) w.r.t. parameter $\theta_\mu$ and noting that the phases of the wavefunctions (\ref{wavefunction}) are equal in magnitude but oppositely signed, we may show that the two terms on the RHS of each equation resulting from the differentiation are equal to each other and thus $\la K_\mp|\pmu|K_\pm\ra=(1/2)\pmu \Delta$. We obtain in this way the following final expressions for the real parts of the various matrix elements in terms of the eigenvector expansion coefficients, $\alpha_\pm,\beta_\pm$:
\ba
\label{realH}
&\Re\Big(\la e_+|\pmu\hrho|e_-\ra\la e_-|\pnu\hrho|e_+\ra\Big)=g_1\la K_+|\pmu|K_+\ra\nn
&\qquad\qquad\qquad\times\la K_+|\pnu|K_+\ra+g_2\pmu\Delta\pnu\Delta;\nn
&\Re \la e_+|\pmu\hrho\pnu\hrho|e_+\ra= h_1^{(+)}\la K_+|\pmu |K_+\ra\la K_+|\pnu|K_+\ra \nn
&+ h_2^{(+)} \pmu\la K_+|\pnu|K_+\ra+h_3^{(+)}\Re \pmu\la K_+|\pnu|K_-\ra\nn
&\qquad\qquad\qquad+ h_4^{(+)}\pmu\Delta\pnu\Delta;\nn
&\Re \la e_-|\pmu\hrho\pnu\hrho|e_-\ra= h_1^{(-)}\la K_+|\pmu |K_+\ra\la K_+|\pnu|K_+\ra\nn
& + h_2^{(-)} \pmu\la K_+|\pnu|K_+\ra+h_3^{(-)}\Re \pmu\la K_+|\pnu|K_-\ra\nn
&\qquad\qquad\qquad+ h_4^{(-)}\pmu\Delta\pnu\Delta,
\end{align}
where
\ba
\label{Hcoeff}
g_1=&-(\alpha_-\beta_+-\alpha_+\beta_-)^2\Delta^2;\nn
g_2=&{1\over 4}\left[(\alpha_-\beta_++\alpha_+\beta_-)+2\Delta(p_+\beta_+\beta_-+p_-\alpha_+\alpha_-)\right]^2;\nn
h_1^{(\pm)}=&p_+^2\alpha_\pm^2+p_-^2\beta_\pm^2+2\alpha_\pm\beta_\pm\Delta (p_+^2+p_-^2+p_+p_-);\nn
h_2^{(\pm)}=&p_+^2(\alpha_\pm+\beta_\pm\Delta)^2+p_-^2(\alpha_\pm\Delta+\beta_\pm)^2;\nn
h_3^{(\pm)}=&2p_+p_-(\alpha_\pm+\beta_\pm\Delta)(\alpha_\pm\Delta+\beta_\pm);\nn
h_4^{(\pm)}=&p_-^2\alpha_\pm^2+p_+^2\beta_\pm^2+2p_+p_-(1+\alpha_\pm\beta_\pm\Delta).
\end{align}

In view of expressions (\ref{cccQFI}) and (\ref{realH}), we may express the QFI matrix element (\ref{QFI}) in terms of four nontrivial matrix elements involving the single-source emission states, $|K_\pm\ra$, as
\ba
\label{QFI1}
{\rm H}_{\mu\nu}= &4\left[{h_1^{(+)}\over e_+}+{h_1^{(-)}\over e_-}-g_1\left({1\over e_+e_-}-1\right)\right]\nn
&\qquad\times \la K_+|\pmu |K_+\ra\la K_+|\pnu|K_+\ra\nn
+&4\left({h_2^{(+)}\over e_+}+{h_2^{(-)}\over e_-}\right)\pmu\la K_+|\pnu|K_+\ra\nn
+&4\left({h_3^{(+)}\over e_+}+{h_3^{(-)}\over e_-}\right)\Re \pmu\la K_+|\pnu|K_-\ra\nn
+&4\Bigg[{h_4^{(+)}\over e_+}+{h_4^{(-)}\over e_-}-g_2\left({1\over e_+ e_-}-1\right)\nn
&\qquad-{3\Delta^2(1-\delta p^2)^2\over \delta e^2(1-\delta e^2)}\Bigg]\partial_\mu \Delta\partial_\nu \Delta,
\end{align}
in which we used the trace norm condition (\ref{eigensum}) to simplify the multipliers of $g_1$ and $g_2$.

The most remarkable fact about the various coefficients in Eq.~(\ref{QFI1}) is that when they are evaluated using expressions (\ref{eigenvalues_coeff}) for $\alpha_\pm,\beta_\pm$, the last two vanish identically and the first two evaluate to be equal to 4, independent of $\Delta$, $p_\pm$, and the separation vector $\bl$, {\it i.e.,}
\be
\label{QFI2}
{\rm H}_{\mu\nu}= 4(\la K_+|\pmu |K_+\ra\la K_+|\pnu|K_+\ra+\pmu\la K_+|\pnu|K_+\ra),
\ee
as shown in Appendix A. Indeed, this means that the QFI matrix elements become identical to those calculated in Ref.~\cite{YuPrasad18} for the problem of a pair of equally bright point sources. The independence of QFI on the relative brightness of the two sources and their separation supports our previous conjecture \cite{YuPrasad18} that the determination of the separation of two point sources when their geometric center is known and fixed {\it a priori} (here at the coordinate origin) reduces fundamentally to a photon localization problem, independent of the nature, brightness, or locations of the point sources. This amounts, in effect, to the irrelevance of which source emits the photon. Rather, the only essential property of relevance to the estimation of pair separation w.r.t. a fixed geometric center is that the photon carrying the information about the sources was in fact emitted and observed by the measuring device.

Expression (\ref{QFI2}) for QFI, as we saw in Ref.~\cite{YuPrasad18}, may be expressed in terms of derivatives of the phase function, $\Psi$, and phase constant, $\phi$. In view of the wave function (\ref{wavefunction}), we may write 
\ba
\label{Dwave}
\la K_+|\pmu|K_+\ra=& i\la\pmu\phi-\pmu\Psi\ra; \nn
\pmu\la K_+|\pnu|K_+\ra=& \la(\pmu\phi-\pmu\Psi)(\pnu\phi-\pnu\Psi)\ra,
\end{align}
in which the angular brackets on the RHS denote pupil-plane averages of the quantity they enclose, 
\be
\label{Pavg}
\la f(\bu)\ra \defeq \int d^2u\, |P(\bu)|^2 f(\bu),
\ee
for any function $f(\bu)$ of the pupil coordinates. When expressions (\ref{Dwave}) are substituted into the RHS of Eq.~(\ref{QFI2}), we can see that the derivatives of the phase constant, namely $\pmu\phi,\pnu\phi$, which do not depend on $\bu$ and can thus be pulled out of such pupil averages, cancel out exactly, reducing QFI to the simple form,
\be
\label{QFI3}
{\rm H}_{\mu\nu}= 4(\la \pmu\Psi\, \pnu\Psi\ra-\la \pmu\Psi\ra\la\pnu\Psi\ra).
\ee
For a clear circular aperture, we may evaluate these averages further and thus obtain \cite{YuPrasad18} the following diagonal form of the QFI matrix:
\be
\label{QFI4}
{\rm \bH}=\left(
\begin{array}{lll}
4\pi^2&0&0\\
0&4\pi^2&0\\
0&0&{\pi^2\over 3}
\end{array}
\right) .
\ee

\section{Quantum Limited Pair Separation for Known Intensity Centroid and Relative Brightness} 
\label{sec:QFI_intensity_centroid}
A more realistic and experimentally meaningful problem involving the separation of a closely spaced source pair is that in which the intensity centroid, rather than the geometric center, of the pair is known to be at a fixed non-random location. Such centroid can be well localized experimentally by finding the centroid of the photon counts recorded by a simple spatial image intensity based array sensor on which the pair is imaged. The intensity centroid, by its very definition, is closer to the brighter source, as for a bright star around which a faintly illuminated planet might be revolving. For the same single-photon DO of the previous section, given by Eq.~(\ref{rho}), the intensity centroid of the pair of sources located at $\br_\pm$ is at $\bR$ given by 
\be
\label{bR}
\bR=p_+\br_++p_-\br_-,
\ee
which implies the following location vectors for the two sources relative to $\vR$:
\be
\label{relvec}
\brho_\pm\defeq \br_\pm-\bR = \pm p_\mp \br,
\ee
in which 
\be
\label{sepvec}
\br= \br_+-\br_-,
\ee
is the separation vector of the two sources. Since the intensity centroid location, $\bR$, can be assumed to be known well in advance, as we just noted, we may pick it to be the origin of coordinates, relative to which $\brho_\pm$ represent the individual source location vectors that are simply proportional, via Eq.~(\ref{relvec}), to the pair separation vector, $\br$, which is to be estimated from source-pair photon measurements. 

The single-photon wavefunctions for single-source emission, namely $\la \bu|K_\pm\ra$, may be expressed in terms of these relative source position vectors as
\be
\label{Cwavefunction}
 \la \bu|K_\pm\ra= \exp(\pm ip_\mp\phi)\, P(\bu)\, \exp[\mp i p_\mp\Psi(\bu;\br)],
\ee
with the phase function $\Psi(\bu;\br)$ being the same as that defined in Eq.~(\ref{pairphase}) with $\bl$ replaced by separation vector $\br$. The choice of the phase constant, $\phi$, is such that  the overlap integral between the two wavefunctions, 
\be
\label{CDelta}
\Delta=\la K_+|K_-\ra =\exp(-i\phi)\int d^2u\, |P(\bu)|^2 \exp[i\Psi(\bu;\br)],
\ee
in which we used the probability normalization, $p_++p_-=1$, becomes not only independent of $p_\pm$ but also real and positive. For this to happen, $\phi$ must be chosen to be the phase of the complex integral on the RHS. In view of the form of the wave function (\ref{Cwavefunction}), we also note the following important correspondences:
\ba
\label{pm}
\la K_-|\partial_\mu|K_-\ra=&-{p_+\over p_-}\la K_+|\partial_\mu|K_+\ra;\nn
\la K_+|\partial_\mu|K_-\ra=&\pmu\la K_-|K_+\ra=p_+\pmu\Delta;\nn
\la K_-|\pmu|K_+\ra=&\pmu\la K_+|K_-\ra=p_-\pmu\Delta.
\end{align}
A straightforward implementation of these correspondences modifies expressions (\ref{realH}) and (\ref{Hcoeff}) into the form
\ba
\label{CrealH}
&\Re\Big(\la e_+|\pmu\hrho|e_-\ra\la e_-|\pnu\hrho|e_+\ra\Big)=\tilde g_1\la K_+|\pmu|K_+\ra\nn
&\qquad\qquad\qquad\times\la K_+|\pnu|K_+\ra+\tilde g_2\pmu\Delta\pnu\Delta;\nn
&\Re \la e_+|\pmu\hrho\pnu\hrho|e_+\ra= \th_1^{(+)}\la K_+|\pmu |K_+\ra\la K_+|\pnu|K_+\ra \nn
&\qquad+ \th_2^{(+)} \pmu\la K_+|\pnu|K_+\ra+\th_3^{(+)}\Re \pmu\la K_+|\pnu|K_-\ra\nn
&\qquad\qquad\qquad+ \th_4^{(+)}\pmu\Delta\pnu\Delta;\nn
&\Re \la e_-|\pmu\hrho\pnu\hrho|e_-\ra= \th_1^{(-)}\la K_+|\pmu |K_+\ra\la K_+|\pnu|K_+\ra \nn
&\qquad+ \th_2^{(-)} \pmu\la K_+|\pnu|K_+\ra+\th_3^{(-)}\Re \pmu\la K_+|\pnu|K_-\ra\nn
&\qquad\qquad\qquad+ \th_4^{(-)}\pmu\Delta\pnu\Delta,
\end{align}
where
\ba
\label{CHcoeff}
\tilde g_1=&-4p_+^2(\alpha_-\beta_+-\alpha_+\beta_-)^2\Delta^2;\nn
\tg_2=&4p_+^2p_-^2\left[(\alpha_-\beta_++\alpha_+\beta_-)+\Delta(\beta_+\beta_-+\alpha_+\alpha_-)\right]^2;\nn
\th_1^{(\pm)}=&p_+^2(\alpha_\pm^2+\beta_\pm^2+6\alpha_\pm\beta_\pm\Delta);\nn
\th_2^{(\pm)}=&p_+^2\left[(\alpha_\pm+\beta_\pm\Delta)^2+(\alpha_\pm\Delta+\beta_\pm)^2\right];\nn
\th_3^{(\pm)}=&2p_+p_-(\alpha_\pm+\beta_\pm\Delta)(\alpha_\pm\Delta+\beta_\pm);\nn
\th_4^{(\pm)}=&p_+^2p_-^2\left[\alpha_\pm^2+\beta_\pm^2+2(1+\alpha_\pm\beta_\pm\Delta)\right].
\end{align}
These new expressions for the coefficients modify the final form of the QFI matrix elements presented earlier in Eq.~(\ref{QFI1}) to the value,
\ba
\label{CQFI1}
\tilde{\rm H}_{\mu\nu}= &4\left[{\th_1^{(+)}\over e_+}+{\th_1^{(-)}\over e_-}-\tg_1\left({1\over e_+e_-}-1\right)\right]\nn
&\qquad\qquad\qquad\times
\la K_+|\pmu |K_+\ra\la K_+|\pnu|K_+\ra\nn
+&4\left({\th_2^{(+)}\over e_+}+{\th_2^{(-)}\over e_-}\right)\pmu\la K_+|\pnu|K_+\ra\nn
+&4\left({\th_3^{(+)}\over e_+}+{\th_3^{(-)}\over e_-}\right)\Re \pmu\la K_+|\pnu|K_-\ra\nn
+&4\Bigg[{\th_4^{(+)}\over e_+}+{\th_4^{(-)}\over e_-}-\tg_2\left({1\over e_+ e_-}-1\right)\nn
&\qquad\qquad\qquad-{3\Delta^2(1-\delta p^2)^2\over \delta e^2(1-\delta e^2)}\Bigg]\partial_\mu \Delta\partial_\nu \Delta.
\end{align}
Since $\th^{(\pm)}_3=h_3^{(\pm)}=0$, the third term in expression (\ref{CQFI1}) is identically zero. The coefficients of the other three terms no longer evaluate to forms that are independent of the pair separation vector $\br$ or the relative intensities of the two sources, as measured by the probability difference, $\delta p$. Their dependence on $\br$ arises through the involvement of the state overlap function, $\Delta$, and its partial derivatives.

In view of the eigenstate normalization condition (\ref{norm3}) and the vanishing of $\th_3^{(\pm)}$, we can immediately simplify the expressions for the coefficients $\th_1^{(\pm)}$, $\th^{(\pm)}_2$, and $\th_4^{(\pm)}$ given by Eq.~(\ref{CHcoeff}) to the form,
\ba
\label{CHcoeff1}
\th_1^{(\pm)}=&p_+^2(1+4\alpha_\pm\beta_\pm\Delta);\nn
\th_2^{(\pm)}=&p_+^2(1+\Delta)^2(\alpha_\pm+\beta_\pm)^2;\nn
\th_4^{(\pm)}=&3p_+^2p_-^2.
\end{align}

Using the fact that $\th_3^{(\pm)}=0$ and substituting the expressions for $\tg_1$ and $\tg_2$, given by Eq.~(\ref{CHcoeff}), and expressions (\ref{CHcoeff1}) into Eq.~(\ref{CQFI1}), we may express the coefficients of the latter in terms of the eigenvector expansion coefficients, $\alpha_\pm$ and $\beta_\pm$, which we may evaluate by using their values given by Eq.~(\ref{eigenvalues_coeff}). The final expression for the QFI matrix elements thus obtained is the following:
\ba
\label{CQFI2}
\tilde{\rm H}_{\mu\nu}= &4\left({1+\delta p\over 1-\delta p}\right)\Bigg[\left(1+{\Delta^2\delta p^2\over 1-\Delta^2}\right)
\la K_+|\pmu |K_+\ra\nn
&\times \la K_+|\pnu|K_+\ra
+\pmu\la K_+|\pnu|K_+\ra\nn
&-\delta p^2(1-\delta p^2)\partial_\mu \Delta\,\partial_\nu \Delta\Bigg].
\end{align}
Note that, as expected, for equal-brightness sources for which $\delta p=0$, expression (\ref{CQFI2}) reduces to the simpler form (\ref{QFI2}) since the geometric and intensity centroids coincide in this case.

Expression (\ref{CQFI2}) for QFI for estimating the 3D pair separation vector when the pair centroid is well localized may be expressed in terms of the derivatives of the phase function $\Psi$, as we did in the previous section. Note, however, that because of unequal coefficients of the first two terms in this expression, the derivatives of the phase constant $\phi$ no longer drop out from the final result obtained in this way. In view of form (\ref{Cwavefunction}), the analogs of Eqs.~(\ref{Dwave}) are the following:
\ba
\label{CDwave}
\la K_+|\pmu|K_+\ra=& ip_-\la\pmu\phi-\pmu\Psi\ra; \nn
\pmu\la K_+|\pnu|K_+\ra=& p_-^2\la(\pmu\phi-\pmu\Psi)(\pnu\phi-\pnu\Psi)\ra,
\end{align}
which when substituted into Eq.~(\ref{CQFI2}) yield the following expression for QFI matrix elements in terms of such derivatives:
\ba
\label{CQFI3}
&\tilde{\rm H}_{\mu\nu}= \left(1-\delta p^2\right)(\la \pmu\Psi\, \pnu\Psi\ra-\la \pmu\Psi\ra\la\pnu\Psi\ra)\nn
-&\delta p^2(1-\delta p^2)\Bigg[{\Delta^2\over 1- \Delta^2}(\la \pmu\Psi\ra-\pmu\phi)(\la\pnu\Psi\ra-\pnu\phi)\nn
&\qquad\qquad\qquad+\partial_\mu \Delta\,\partial_\nu \Delta\Bigg].
\end{align}
By differentiating the equality,
\be
\label{dphi}
\Delta \exp(i\phi)=\la \exp(i\Psi)\ra, 
\ee
which follows from Eq.~(\ref{CDelta}), then dividing the result by $\Delta\exp(i\phi)$, and subsequently evaluating the real and imaginary part of the resulting expression, we may calculate $\pmu\Delta$ and $\pmu\phi$ as
\ba
\label{DDeltaphi}
\pmu\Delta =&-\Delta\, \Im\left[{\la\pmu\Psi\exp(i\Psi)\ra\over \la\exp(i\Psi)\ra}\right];\nn
\pmu \phi =&\Re \left[{\la\pmu\Psi\exp(i\Psi)\ra\over \la\exp(i\Psi)\ra}\right]. 
\end{align}
Use of these equalities and the equality, $\Delta=|\la\exp(i\Psi)\ra|$, in Eq.~(\ref{CQFI3}) turns it into an expression composed of pupil-plane averages of form (\ref{Pavg}) of quantities involving only the phase function $\Psi$, given by Eq.~(\ref{pairphase}), and its derivatives. 
 
\section{Numerical Results for a Circular Aperture}

The results we have derived in the previous section apply to arbitrarily-shaped aperture with any pupil function, $P(\bu)$, with the only constraint on the latter being the normalization condition (\ref{norm}). The 3D pair-separation coordinates, $l_x,l_y,l_z$, are the corresponding physical coordinates, $(x,y,z)$, divided by certain lateral and axial diffraction-dependent scales, namely $\lambda z_I/R$ and $\lambda z_I^2/R^2$, in which $\lambda$ is the optical wavelength, $R$ represents the characteristic size of the exit pupil, and $z_I$ is the image-plane distance from the exit pupil. Specifically, 
\be
\label{ScaledCoords}
(l_x,l_y)={(x,y)\over \lambda z_I/R}, \  l_z={z\over \lambda z_I^2/R^2}.
\ee
For a thin-lens, circular-aperture imager, for which the pupil and aperture planes coincide, $R$ is simply its radius.

We limit our numerically obtained results to the case of such a circular aperture that is furthermore clear, for which $\bu=\brho/R$ and $|P(\bu)|^2$ is simply $1/\pi$ for $\bu|<1$ and 0 otherwise. For this pupil geometry, we see easily from its inversion symmetry that
\be
\label{dPsi}
\la\partial_x\Psi\ra\! \sim\! \la u_x\ra=0; \  \la\partial_y\Psi\ra\! \sim\! \la u_y\ra=0;\ 
\la\partial_z\Psi\ra = \pi\la u^2\ra = {\pi\over 2}.
\ee
Similar integrations over the pupil yield the following pupil averages:
\be
\label{dPsi_dPsi}
\la \pmu\Psi\pnu\Psi\ra =\left\{
\begin{array}{ll}
\pi^2, &\mu=\nu=x,y\\
\pi^2/ 3, &\mu=\nu=z\\
0, &{\rm otherwise}.
\end{array}
\right.
\ee 
Results (\ref{dPsi}) and (\ref{dPsi_dPsi}) imply the following diagonal form for the difference of averages on the first line of Eq.~(\ref{CQFI3}):
\be
\label{CQFI0}
\la \pmu\Psi\pnu\Psi\ra -\la\pmu\Psi\ra\la\pnu\Psi\ra=\left\{
\begin{array}{ll}
\pi^2 &\mu=\nu=x,y\nn
\pi^2/12 &\mu=\nu=z\nn
0 &{\rm otherwise}
\end{array}
\right.,
\ee
which thus contribute the terms $(1-\delta p^2)\pi^2(1,1,1/12)$ to the three diagonal elements of the QFI. 

The other terms in Eq.~(\ref{CQFI3}), which only contribute for an unequal-brightness pair $(\delta p\neq 0)$, may be evaluated, as Eq.~(\ref{DDeltaphi}) shows, by calculating the three pupil-plane averages,
\ba
\label{R0123plane}
&\Delta=|\la\exp(i\Psi)\ra|;\nn
&\la \px\Psi\exp(i\Psi)\ra\!=\!2\pi \la u\cos\phi_u \exp(i2\pi\bu\cdot\blperp)\exp(i\pi u^2 l_z)\ra;\nn
&\la \py\Psi\exp(i\Psi)\ra\!=\!2\pi \la u\sin\phi_u \exp(i2\pi\bu\cdot\blperp)\exp(i\pi u^2 l_z)\ra;\nn
&\la \pz\Psi\exp(i\Psi)\ra\!=\!\pi \la u^2 \exp(i2\pi\bu\cdot\blperp)\exp(i\pi u^2 l_z)\ra.
\end{align}
These averages, which are pupil-plane integrals over the circular aperture, may be simplified by use of the angular-integral identities,
\ba
\label{BesselIdentities}
\oint d\phi_u \exp(i2\pi \bu\cdot\blperp)=&2\pi J_0(2\pi u l_\perp),\nn
\oint d\phi_u \cos\phi_u \exp(i2\pi \bu\cdot\blperp)=&i2\pi \cos\phi_l J_1(2\pi u l_\perp),\nn
\oint d\phi_u \sin\phi_u \exp(i2\pi \bu\cdot\blperp)=&i2\pi \sin\phi_l J_1(2\pi u l_\perp),
\end{align}
in terms of Bessel functions $J_0$ and $J_1$,
to the following radial integrals:
\ba
\label{R0123radial}
\la\exp(i\Psi)\ra=&2\int_0^1\!\! du \, u J_0(2\pi ul_\perp)\,\exp(i\pi u^2l_z),\nn
\la \px\Psi\exp(i\Psi)\ra\!=&4i\pi  \cos\phi_l \int_0^1\!\!\!du \, u^2J_1(2\pi u\,l_\perp)\exp(i\pi u^2 l_z),\nn
\la \py\Psi\exp(i\Psi)\ra\!=&4i\pi  \sin\phi_l \int_0^1\!\!\!du \, u^2 J_1(2\pi u\,l_\perp)\exp(i\pi u^2 l_z),\nn
\la \pz\Psi\exp(i\Psi)\ra\!=&2\pi \int_0^1\!\!\!du\,u^3 J_0(2\pi u\,l_\perp)\exp(i\pi u^2 l_z).
\end{align}
We numerically evaluated these radial integrals in Matlab using its built-in {\it integral} code, from which by taking the real and imaginary parts of the ratios involved in Eq.~(\ref{DDeltaphi}) we were able to calculate all of the terms occurring on the second line of Eq.~(\ref{CQFI3}). After evaluating all of the QFI matrix elements in this way, we were able to invert this $3\times 3$ matrix to calculate the QCRBs as the diagonal elements of this inverse matrix. These QCRBs serve to provide the lowest possible bounds on the variance of unbiased estimation of the three separation coordinates of the pair per photon. Dividing the QCRB per photon for each coordinate by the mean photon number involved in the measurement then generates the overall QCRB corresponding to the estimation of that coordinate.  

In Figs.~1, we display, using a surface plot, QCRB per photon for the estimation of the $x$ coordinate of the transverse separation as a function of $(l_x,l_y)$ for three different values of $\delta p^2$, namely 0, 0.75, and 0.95, corresponding to the ratio of source brightnesses taking the values 1:1, 14:1, and 78:1, respectively. The three different sub-figures refer to $l_z$ taking values 0 (in-focus), $\pm 1$, and $\pm 2$ units, respectively. 

Figures 2, on the other hand, capture the numerically evaluated values of QCRB for the estimation of the axial separation coordinate, $l_z$, for the same conditions as in Figs.~1. Apart from a factor of 4 owing to a difference by a factor 2 in the definitions of the separation vector between Ref.~\cite{YuPrasad18} and the present section, QCRB for the equally bright source pair, corresponding to $\delta p=0$, is a constant equal to $1/\pi^2$ for estimating each of the two transverse separations and $12/\pi^2$ for estimating the axial separation, independent of the actual value of the separation vector. These results are represented by the bottom flat surface in each sub-figure. 

As the two source brightness values begin to differ from each other, QCRB with respect to the estimation of each of the tranverse and axial separation coordinates shows significant variations at small transverse separations. These variations tend to be characterized by peaks and valleys over a considerable range of such short separations. They result from diffraction generated blurring of the wavefronts passing through a finite aperture, which is represented by the oscillatory behavior of Bessel functions of low order. When one source is considerably fainter than the other, then the two sources would be harder to separate from each other if the former were to lie in a minimum of the transverse diffraction pattern of the latter. The same considerations apply to axial separations as well since there is a diffraction-generated oscillatory spreading in the axial dimension as well. 

These oscillatory peaks and valleys are accentuated by increasing asymmetry of source brightness levels, particularly at short axial and lateral separations, as a comparison of the three surfaces in each figure shows.  As the sources get well separated in their transverse dimensions, QCRB becomes asymptotically independent of separation as the sources then lie well outside the diffractive footprint of each source and can be well distinguished from each other. The asymptotic values of QCRB are, however, larger the larger the brightness asymmetry, since locating the fainter source - and thus estimating its separation from the brighter source - becomes noisier the larger such asymmetry. 

\begin{figure}[!ht]
     \subfloat[\label{subfig_1a}]{%
       \includegraphics[width=0.45\textwidth]{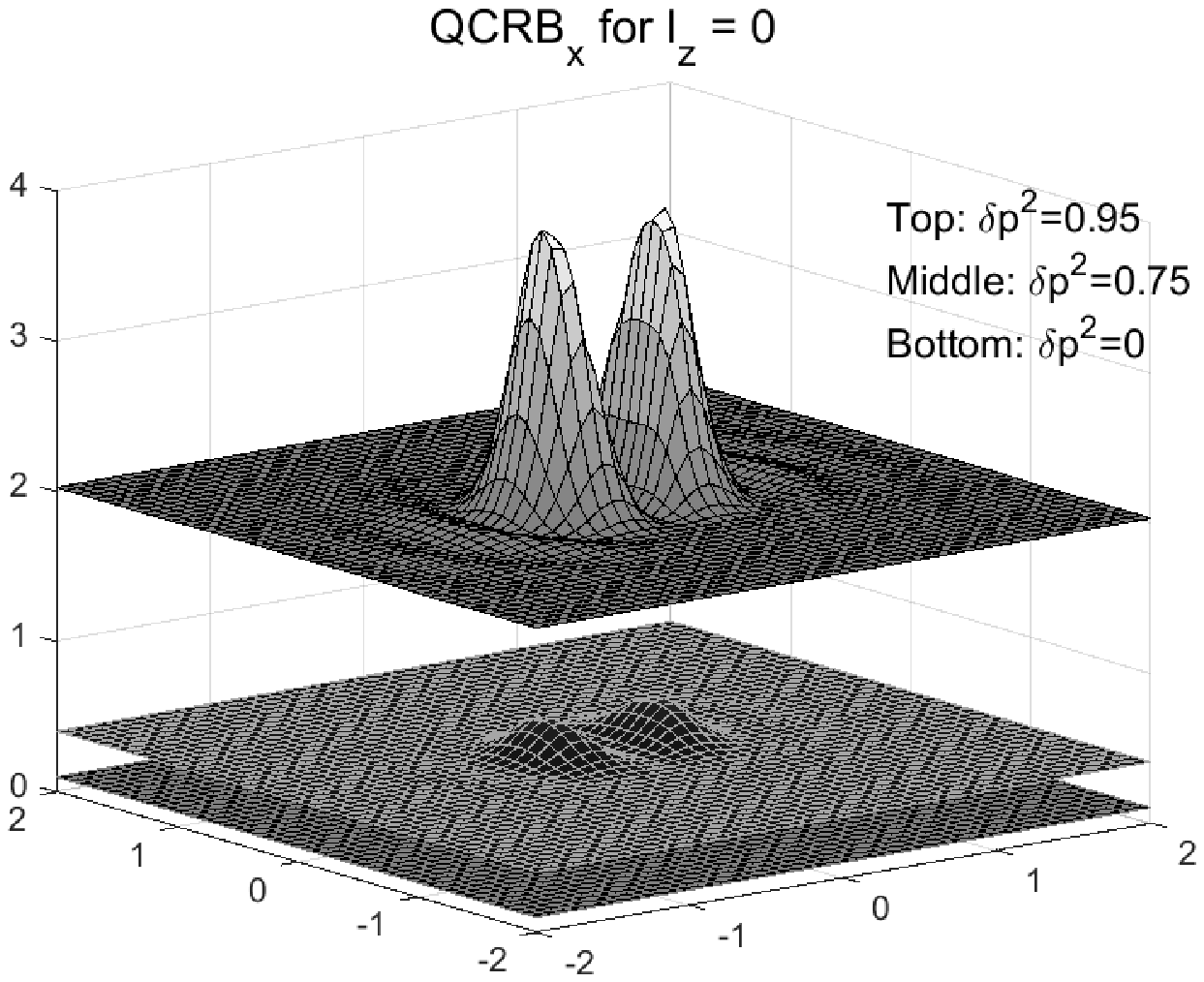}
     }
     \hfill
     \subfloat[\label{subfig_1b}]{%
       \includegraphics[width=0.45\textwidth]{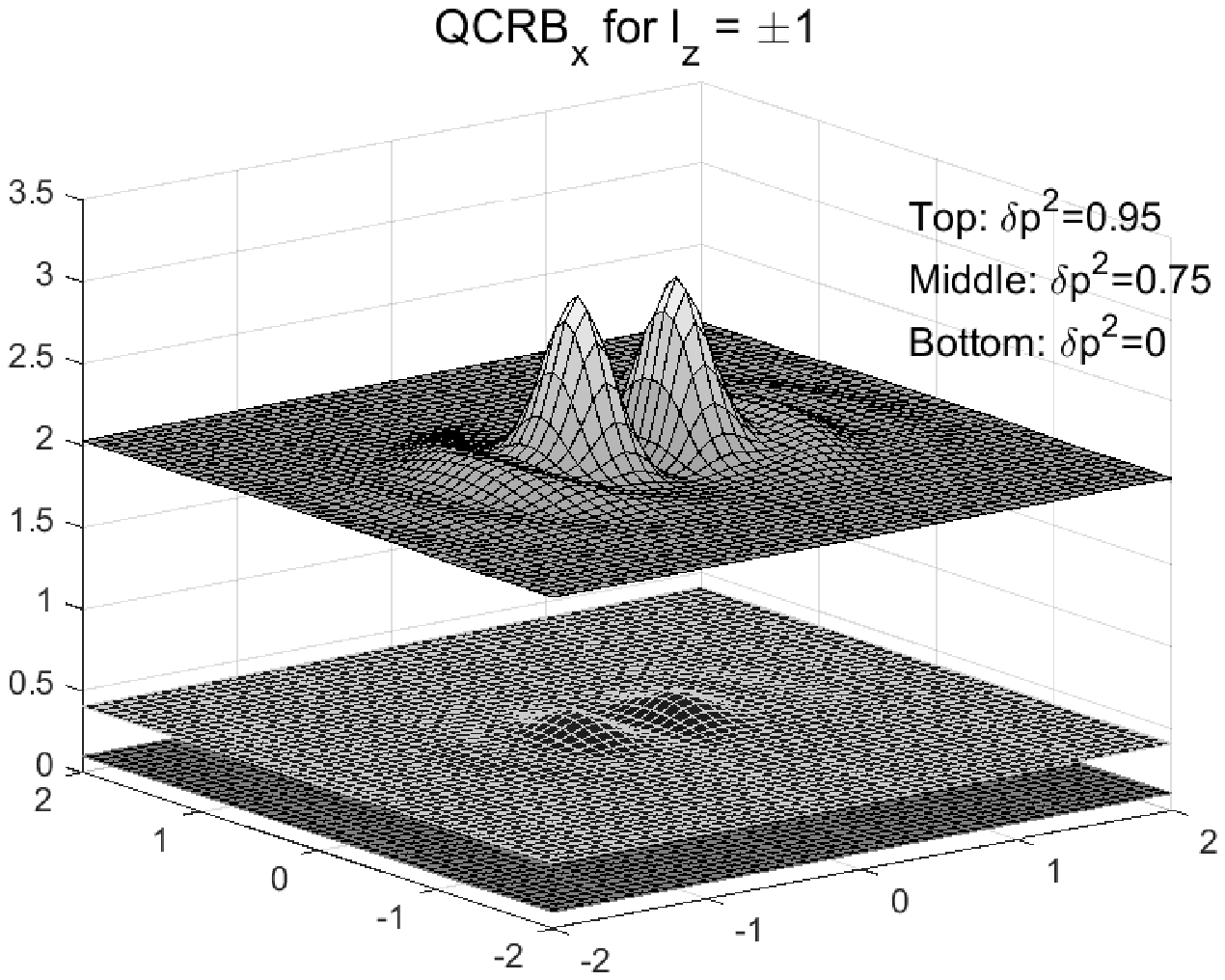}
     }
     \hfill
     \subfloat[\label{subfig_1c}]{%
       \includegraphics[width=0.45\textwidth]{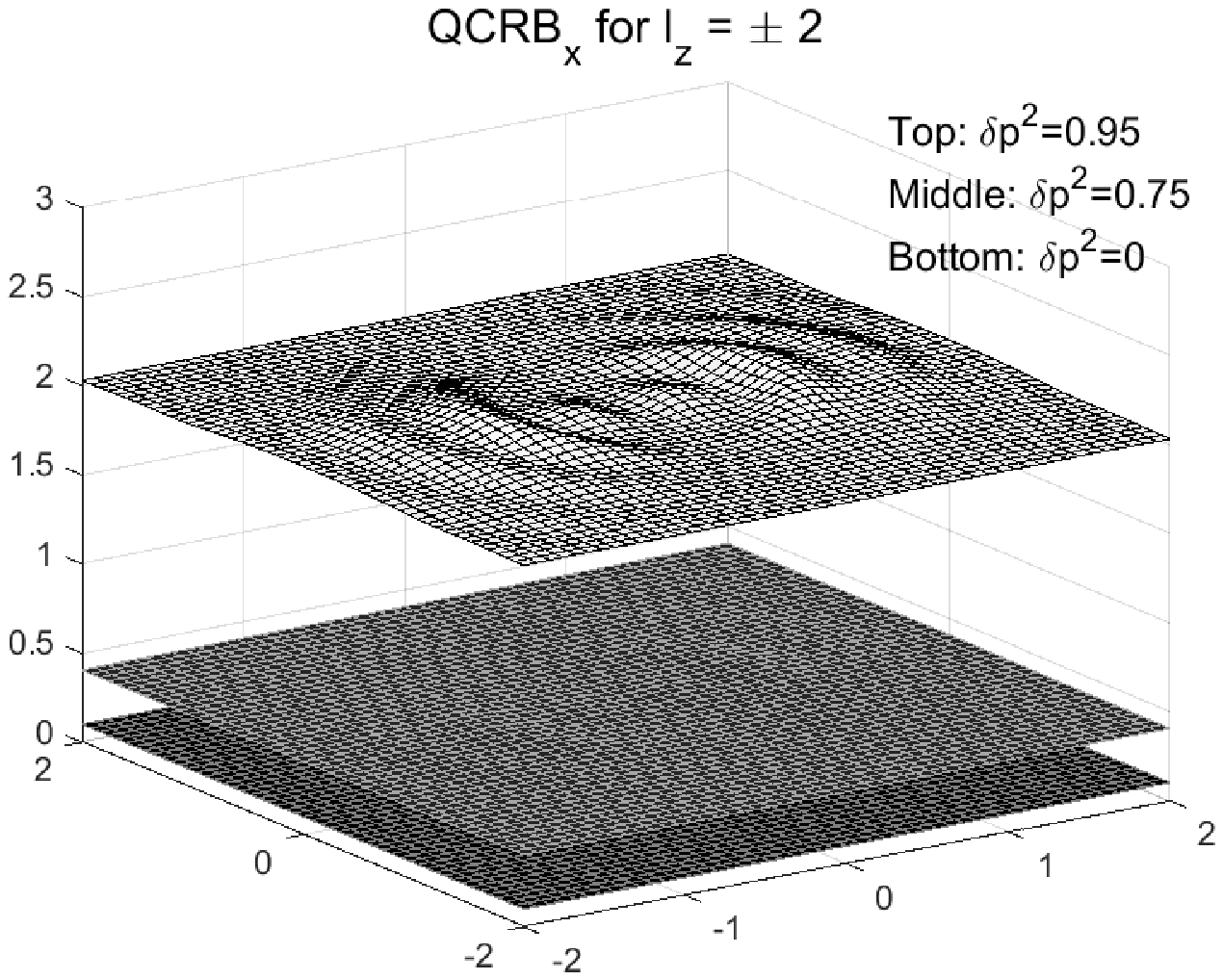}
     }
     \caption{Surface plots of QCRB for $l_x$ as a function of $(l_x,l_y)$ for (a) $l_z=0$, (b) $l_z=\pm 1$, (c) $l_z=\pm 2$, and for three different values of $\delta p^2$, namely 0 (blue surface), 0.45 (red surface), and 0.95 (green surface)}
     \label{fig:Fig1}
   \end{figure}

\begin{figure}[!ht]
     \subfloat[\label{subfig_2a}]{%
       \includegraphics[width=0.45\textwidth]{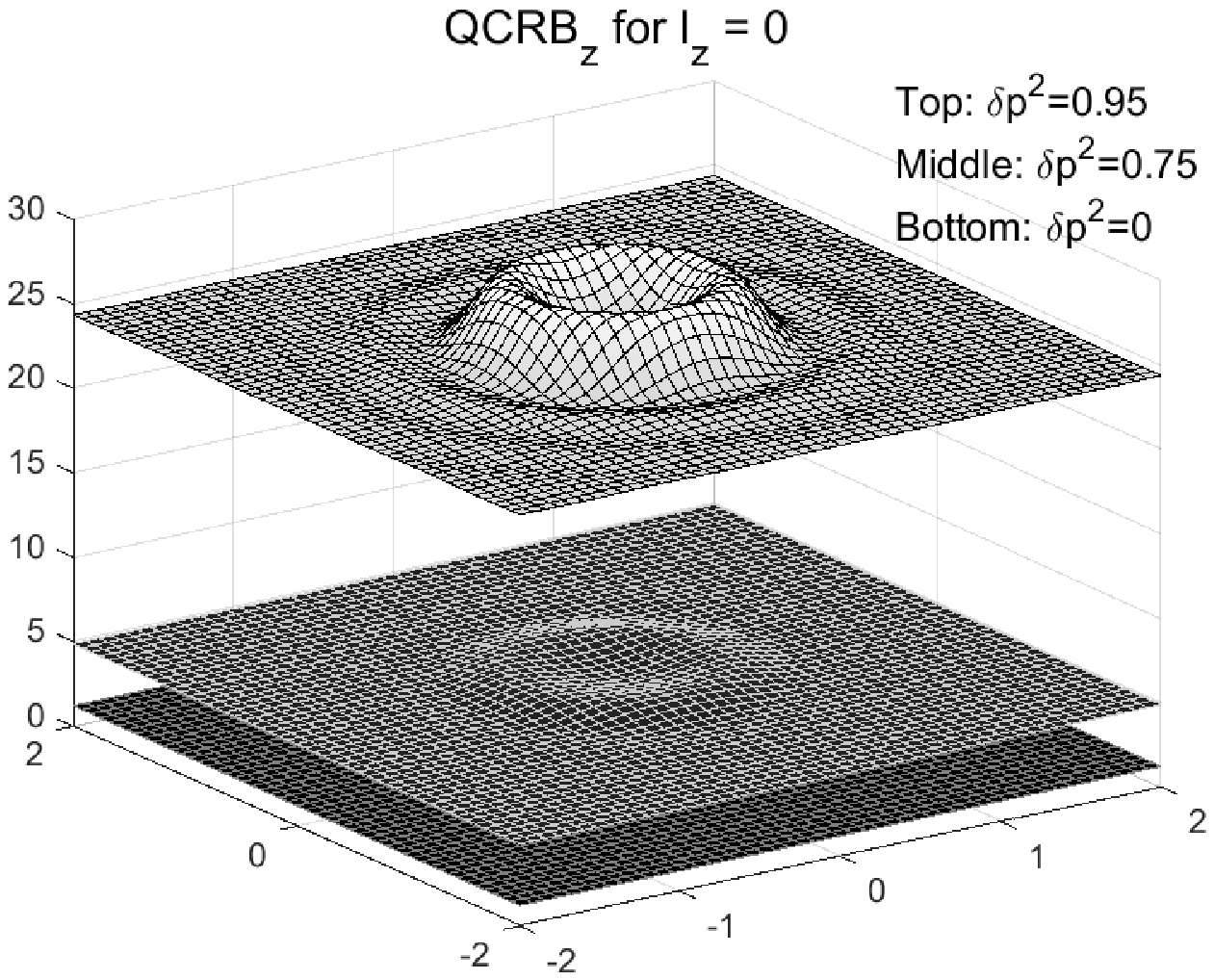}
     }
     \hfill
     \subfloat[\label{subfig_2b}]{%
       \includegraphics[width=0.45\textwidth]{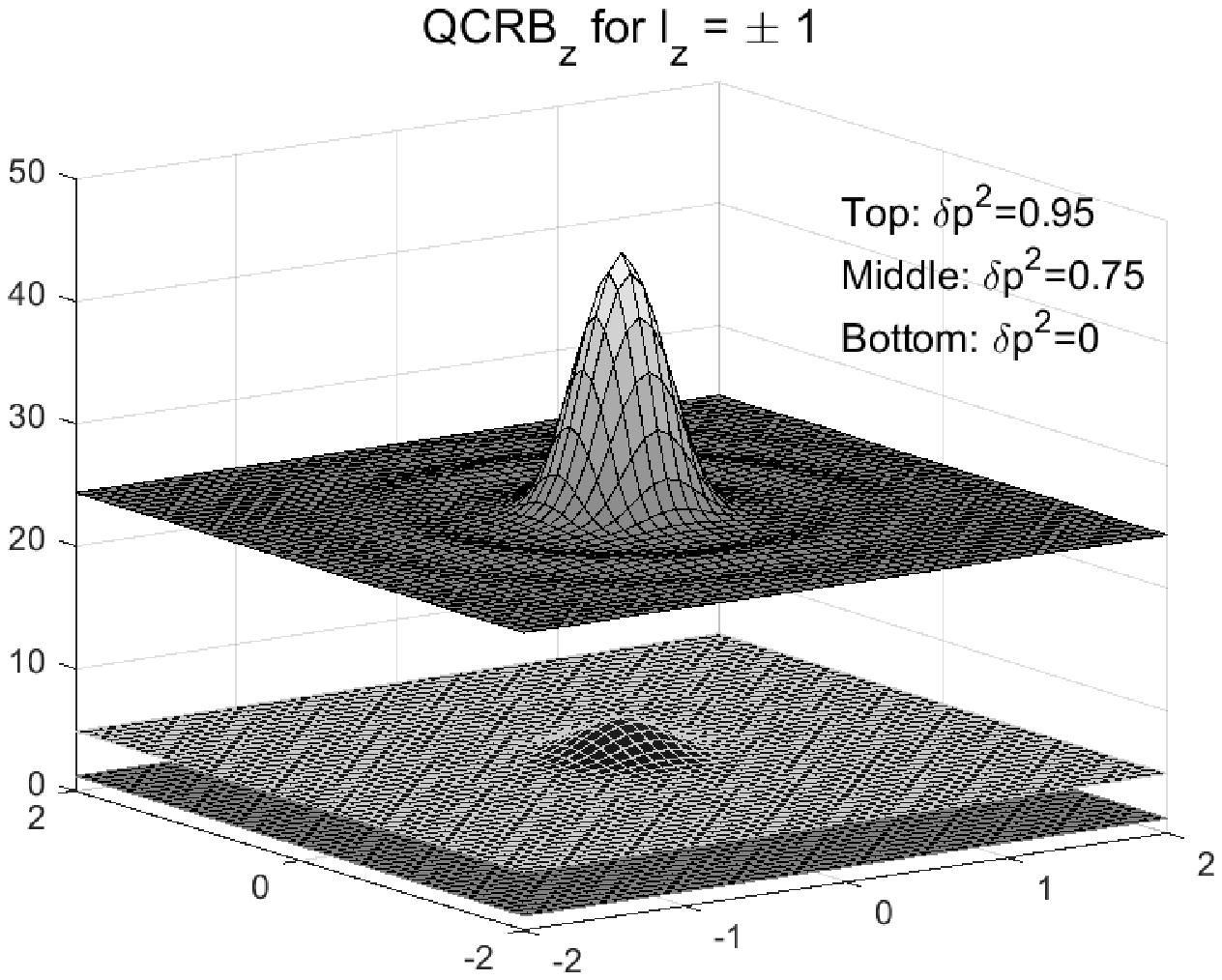}
     }
     \hfill
     \subfloat[\label{subfig_2c}]{%
       \includegraphics[width=0.45\textwidth]{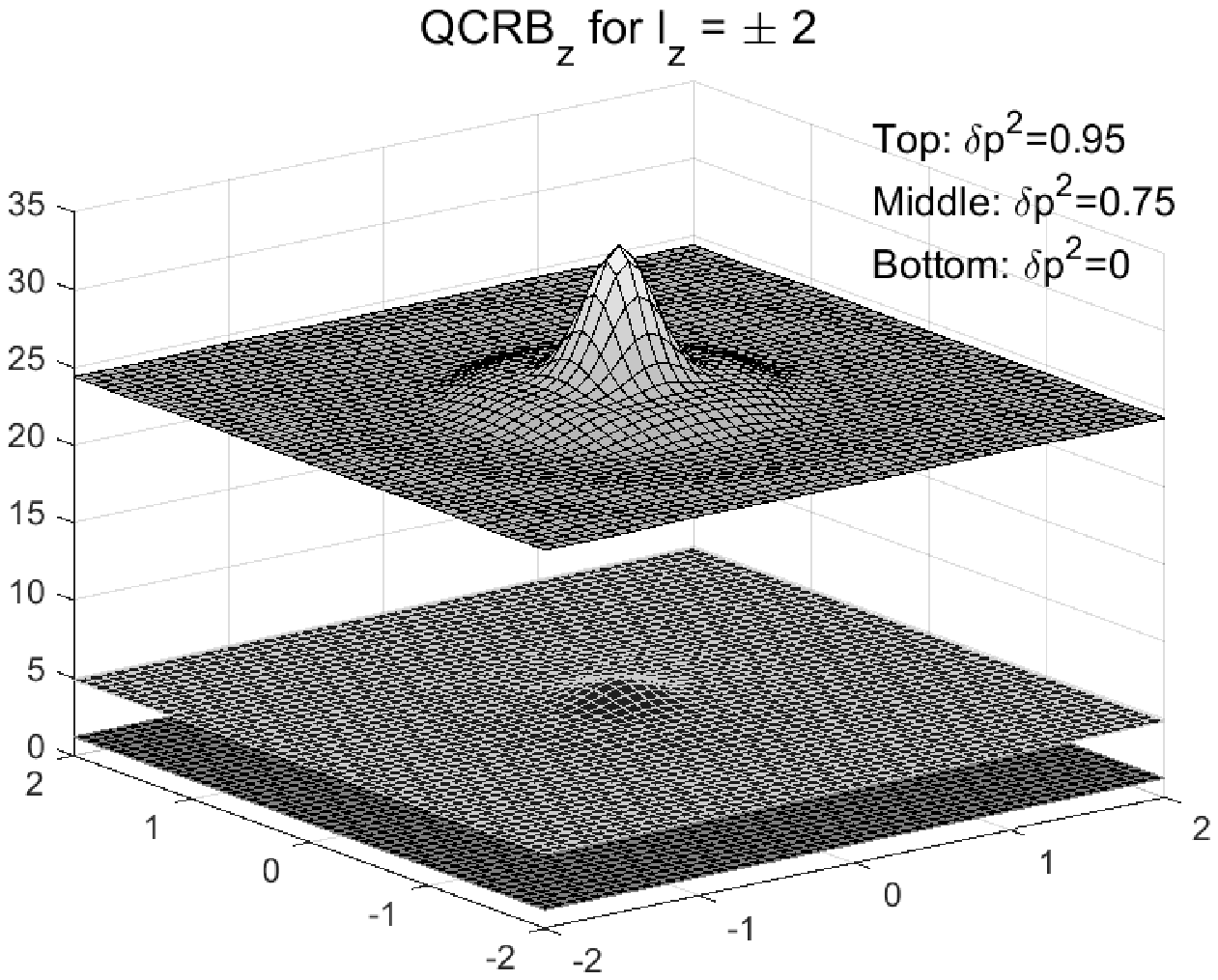}
     }
     \caption{Surface plots of QCRB for $l_z$ for the same parameter values as in FIG.~1}
     \label{fig:Fig2}
   \end{figure}

The values of QCRB for the estimation of the second transverse separation coordinate, $l_y$, are identical to those in Figs.~1 when they are rotated by $\pi/2$ because of the full $l_x\leftrightarrow l_y$ symmetry of QCRB obtained by a $\pi/2$ rotation about the $z$ axis, followed by a mirror reflection in one of the transverse axes. As such, we do not plot such QCRB values separately.

We also note that Figs.~1 and 2, corresponding to the estimation of the lateral and axial separation coordinates, respectively, differ in the asymptotic values of the two different QCRBs by a factor of 12 by which the two estimation variances differ from each other. This can be seen from the inverse of the diagonal QFI matrix corresponding to the first line of terms in Eq.~(\ref{CQFI3}). The second line of terms in that equation tend to vanish in the asymptotic limit of large separations. 

\section{Concluding Remarks}
This paper has computed the quantum lower bounds on the variances for estimating the three components of the vector separation of a pair of incoherent point sources of unequal brightness. For the case of fixed intensity centroid of the source pair, the QFI matrix with respect to these three components and its inverse, whose diagonal elements furnish the said lower bounds, are both diagonal and independent of the specific values of the separation parameters only when the sources are equally bright. On the other hand, when the source pair's geometric center is well determined {\it a priori}, then QFI becomes independent of the separation coordinates as well as the brightness ratio for the pair, reducing to the previously obtained simple form \cite{YuPrasad18} for an equally bright pair of sources. The latter result supports the conjecture that for fixed geometric center of the pair, a photon bringing the information about the location of either source to the imager can be interrogated with a fidelity that is independent of which source emitted it, since the two, by the very definition of the geometric center, have equal but oppositely signed vector separation from it. The experimental usefulness of this result is somewhat limited, however, since the more asymmetric the brightnesses of the two sources, the more difficult it would be to locate their geometric center precisely. The intensity centroid, by contrast, can be more readily and precisely located within the standard imaging protocol by computing the centroid of the photon counts observed on an array imaging sensor. 

In Ref.~\cite{YuPrasad18} we demonstrated, by means of a numerical simulation, the use of low-order Zernike modes to saturate QCRB for the case of a symmetrical pair in the limit of small separations. We expect the Zernike projection basis to saturate QCRB regardless of the pair's brightness asymmetry, since the dependence of the phase of the photon wavefunction  on separation, as seen from Eq.~(\ref{Cwavefunction}), is formally invariant under a change of such asymmetry represented by a change of $p_\pm$ away from 1/2.

Future work will enlarge the scope of the problem to include the quantum bound on the fidelity for the estimation of the brightness ratio for the source pair and for separating more than two unequally bright point sources in close vicinity of one another with a known brightness centroid. Such studies could be important as a first step in calculating the quantum limits on 3D superresolution imaging of extended continuous sources.

\section*{Acknowledgments}
The author is honored to recognize the immense contribution that Roy J. Glauber made to his scholarly development as a physicist, particularly in the field of modern quantum optics. It all began with his Ph.D research under Roy's guidance during the period 1978-1983, and then continued over the next three decades of collaborative work. Roy showed much interest in the author's early work in the area of quantum limited pair superresolution, but due to its unfortunate timing the present work is the poorer for having failed to benefit from Roy's sharp insights into quantum physics. The author also acknowledges helpful discussions with Zhixian Yu. The work has been partially supported by a consulting agreement with the Boeing Company.

\end{document}